\documentclass[graybox]{svmult}

\usepackage{mathptmx}       
\usepackage{helvet}         
\usepackage{courier}        
\usepackage{type1cm}        
\usepackage{makeidx}         
\usepackage{graphicx}        
\usepackage{multicol}        
\usepackage[bottom]{footmisc}

\makeindex             

\usepackage{amssymb,amsmath,multirow,rotate,color}
\usepackage{url}
\usepackage[english]{babel}

\newcommand{\etal}{\emph{et al.}}
\newcommand{\avg}[1]{\langle #1 \rangle}

\graphicspath{{pictures/}}

\begin{document}

\title*{Several multiplexes in the same city: The role of socioeconomic differences in urban mobility}
\titlerunning{\textsc{Urban mobility multiplex networks}} 

\author{Laura Lotero, Alessio Cadillo,  Rafael Hurtado and Jes\'us G\'omez-Garde\~nes}
\authorrunning{\textsc{Lotero \etal}}

\institute{%
Laura Lotero \at Departamento de Ciencias de la Computaci\'on y de la Decisi\'on, Universidad Nacional de Colombia, Medell\'in, Colombia, \email{llotero0@unal.edu.co}
\and Alessio Cardillo \at Laboratoire de Biophysique Statistique, \'Ecole Polytechnique F\'ed\'erale de Lausanne (EPFL), CH-1015 Lausanne, Switzerland. Instituto de Biocomputaci\'on y F\'{\i}sica de Sistemas Complejos, Universidad de Zaragoza, E-50018 Zaragoza, Spain, \email{alessio.cardillo@epfl.ch}
\and Rafael Hurtado \at Departamento de F\'{\i}sica, Universidad Nacional de Colombia, Bogot\'a, Colombia, \email{rghurtadoh@unal.edu.co} 
\and Jes\'us G\'omez-Garde\~nes \at Departamento de F\'{\i}sica de la Materia Condensada, Universidad de Zaragoza, E-50009 Zaragoza, Spain.
Instituto de Biocomputaci\'on y F\'{\i}sica de Sistemas Complejos, Universidad de Zaragoza, E-50018 Zaragoza, Spain, \email{gardenes@gmail.com}
} 

\maketitle

\abstract{In this work we analyze the architecture of real urban mobility networks from the multiplex perspective. In particular, based on empirical data about the mobility patterns in the cities of Bogot\'a and Medell\'{\i}n, each city is represented by six multiplex networks, each one representing the origin-destination trips performed by a subset of the population corresponding to a particular socioeconomic status. The nodes of each multiplex are the different urban locations whereas links represent the existence of a trip from one node (origin) to another (destination). On the other hand, the different layers of each multiplex correspond to the different existing transportation modes. By exploiting the characterization of multiplex transportation networks combining different transportation modes, we aim at characterizing the mobility patterns of each subset of the population. Our results show that the socioeconomic characteristics of the population have an extraordinary impact in the layer organization of these multiplex systems.}

\newpage

%
%

\section{Introduction}
\label{sec:intro}

Understanding human mobility patterns have attracted,  for decades, the attention of researchers from many different scientific realms. The first models based on empirical observations date back to the $40$'s and were elaborated by the sociologist Samuel A. Stauffer \cite{stouffer-1940} and the philologist George K. Zipf \cite{zipf-1946}. It is at the end of the last century when, with the consolidation of mathematical frameworks such as the well-known {\em Gravity model} \cite{erlander-1990}, when transportation science, as a realm of operations research, became a discipline on its own.

The advent of the big data era, have spurred the activity on transportation science and provided detailed datasets of real transportation systems. This characterization spans across many scales, from the short-range mobility patterns in urban areas \cite{batty-science-2008,porta-barcelona,porta-bologna} to world wide trips \cite{guimera-pnas-2005}. Remarkably, different degrees of resolution and types of information are nowadays available from the combined use of techniques for data gathering \cite{asgari-arxiv-2013}. From the traditional datasets based on direct surveys \cite{yan-scirep-2013}, allowing to know the purpose of the trip (work/school, leisure, etc), to those large-scale ones gathered by tracking mobile communication systems \cite{gonzalez-nature-2008,wang-scirep-2012} or transport electronic cards \cite{roth-plosone-2011}. This burst of activity have attracted many scientist from theoretical disciplines to contribute to the subject through the formulation of mobility models and mathematical tools aimed at reproducing and characterizing the observed patterns of movement \cite{helbing-pedestrian-2005,bazzani-acs-2007,song-natphys-2010,simini-nature-2012}.

The rapid change in the patterns of human mobility in the last decades, specially in what concerns the decrease in their duration together with the increase of their length, makes its characterization of utmost importance  for many disciplines beyond the traditional scope of transportation science. The most paradigmatic example is the relevance of human mobility in the spread of diseases. The inclusion of the mobility ingredient into epidemic models has allowed to design sophisticated theoretical frameworks aiming at forecasting the onset and duration of pandemics with high time and spatial resolution \cite{eubank-nature-2004,colizza-pnas-2006,kleinberg-nature-2007,balcan-pnas-2009,tizzoni-bmc-2012,poletto-jtb-2013}.

In the last fifteen years, networks science \cite{barabasi-rev-2002,newman-rev-2003,boccaletti-rev-2006} have appeared as the best suited mathematical frameworks to accommodate and characterize the interaction backbone of the very many complex systems captured by big data techniques. In fact, complex networks had been proposed as the natural framework to study spatially embedded systems \cite{barthelemy-rev-2011} and, in particular, mobility networks. In these networks the different origins and destinations are represented as nodes of a graph, whereas the movements between locations are encoded as links connecting them \cite{strano-bycicle-plos-2013}. Recently, thanks to the availability of more detailed information, it has been possible to represent many different types of transportation modes used for the movements within the same area under multilayer networks \cite{de_domenico-prx-2013,boccaletti-rev-2014,kivela-rev-2014}, in which network layer represents a single transportation mode. In this way, each node still represents a particular origin/destination location and it is present in each of the network layers. However, links are represented in a different layer of interaction depending on the kind of transportation mode used for connecting two locations. This particular multilayer network is usually termed as {\em multiplex}. 

In the recent years, different human mobility systems have been addressed under the paradigm of multiplex networks, ranging from urban movements \cite{manlio-pnas-2014} to medium \cite{kurant-prl-2006} and large scale trips \cite{cardillo-epjst-2013,cardillo-scirep-2013}.
Following this approach, here we address the multiplex structure of urban mobility in two different cities: Bogot\'a and Medell\'{\i}n. The novelty of the results presented rely on an additional ingredient of the mobility patterns that, up to our knowledge, has been ignored up to date. This new ingredient is the socioeconomic status {\em (SES)} of the individuals, mainly related to their wealth. Being the composition of many cities in the world highly hierarchical and inhomogeneous in terms of the capital distribution, it is thus relevant to unveil the influence that the different SES have on the mobility patterns. 

To this aim, and considering that another relevant ingredient included in the available data sets is the transportation modes used by the individuals, we analyze the mobility patterns in terms of a multiplex network. In particular, we will analyze six different multiplex networks, each one corresponding to a different SES. Our approach relies on the {\em adiabatic projection} technique, introduced in \cite{cardillo-scirep-2013}, that consist in monitoring how the structural properties of the aggregate network show up as a result of the merging of the layers composing the multiplex. Thanks to this approach, it has been possible to spotlight how {\em segregation} and {\em multimodality} are characteristic of some particular social classes, and to unveil the dominant role played by the middle-class in the utilization of the transportation system as a whole.

The structure of the paper is the following. We will first introduce the datasets used in section~\ref{sec:data} and the adiabatic projection technique together with the topological estimators in which it is used in section~\ref{sec:methods}. Section \ref{sec:results} is devoted to present the results of applying the former technique to the datasets of the cities of  Bogot\'a and Medell\'{\i}n. Finally, in section~\ref{sec:conclusions} we draw some conclusions and future work perspectives. 

%
%

\section{Urban Mobility and Socioeconomic Status}
\label{sec:data}

The mobility data presented and analyzed here are taken from surveys carried out in two major cities of Colombia:  Bogot\'a and Medell\'in. These surveys were originally designed to collect information about travelers and their trips, so to identify traffic patterns and apply the results to urban and transportation planning. In these surveys, each householder is asked about the trips performed the day before the interview, providing with the origin and destination zones, the departure and arrival times, the transportation mode used and the purpose of each trip. In addition, householders are characterized by their socioeconomic characteristics, such as the age, gender, occupation, and the socioeconomic characteristic of their housing, which it is defined as its SES. The survey for the city of Bogot\'a, having a population of about $7$ million of inhabitants, has a sample size of $45446$ people interviewed, reporting $100846$ trips \cite{dataset1}. On the other hand, the survey for the metropolitan area of Medell\'in, with a population of about $3.5$ million people, reports $127849$ trips from $56513$ personal interviews \cite{dataset2}. However, not all the people interviewed made a trip and thus the number of travelers in both cities is smaller (see Table~\ref{tab:datasets}).

\begin{table}[b!]
\centering
\begin{tabular}{|l||c|c|c|c|c|c|c|c|}\hline
  & $P$ & $\avg{A}$ & $Fem (\%)$ & $T_{\text{{\sc TOT}}}$ & $\avg{T}$ & $\avg{n_s}$ & $N$ & $E$ \\\hline\hline
\textbf{Bogot\'a} & 37483 & 33.23 & 54.4 & 100846 & 2.69 & -- & 912 & 24588\\\hline
\textbf{Medellin} & 45496 & 33.34 & 53.4 & 127849 & 1.58 & 1.105273 & 413 & 18442 \\\hline
\end{tabular}
\caption{ Supplementary information about the mobility interviews. From left to right: number of travellers, $P$, their average age, $\avg{A}$, percentage of female subjects, $Fem$, total number of trips recorded, $T_{\text{{\sc TOT}}}$, average number of trips per person, $\avg{T}$, average number of steps per trip, $\avg{n_s}$, number of urban areas (nodes), $N$, number of connections between areas (links), $E$. }
\label{tab:datasets}
\end{table}

\begin{figure}[t!]
\centering
\includegraphics[width=0.88\textwidth]{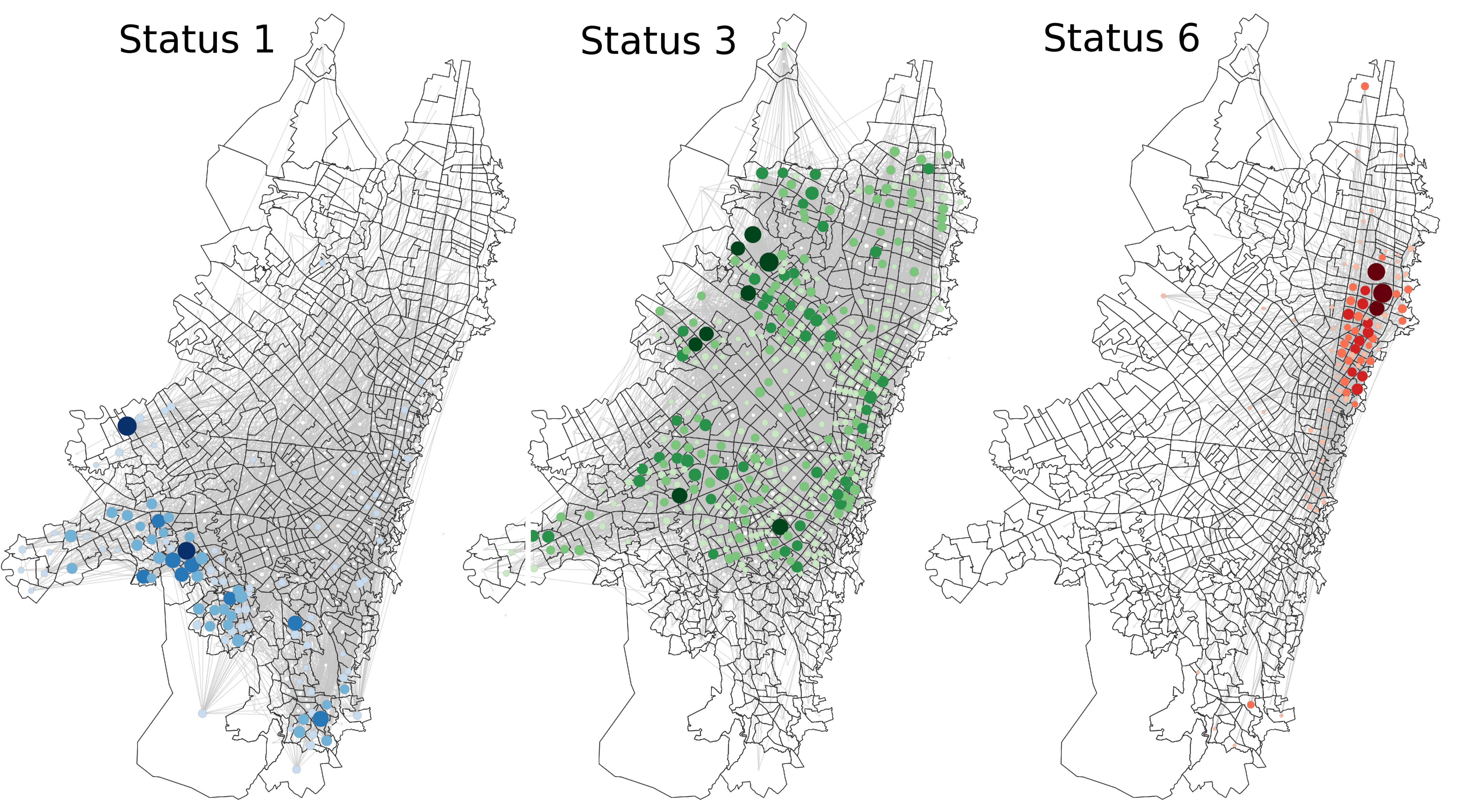}
\caption{Maps of the trips made in the city of Bogot\'a. Each map refers to a particular socioeconomic status, namely (from left to right) $1$, $3$ and $6$. Each node corresponds to a different urban mobility zone, while a link (light gray) between two nodes indicates that a displacement has occurred among them. The size of the nodes indicates the amount of displacements (both from and to) occurring in that zone.}
\label{fig:estratos-sociales}
\end{figure}

In Table \ref{tab:datasets} we briefly show the most relevant information about the population interviewed in both cities. From the network perspective, the mobility graphs derived from these surveys contain $N=912$ nodes (being both origins and destinations) for the city of Bogot\'a,  and $N=413$ for the metropolitan area of Medell\'{\i}n. In this way, two nodes are linked whenever the survey reports the existence of at least one trip between two zones. In addition, we take advantage of the socioeconomic information provided by the surveys, in particular the information about the SES of each individual, being this a good proxy of the population wealth. This categorization of the population into strata is specific of Colombia, and ranges from status $1$ for the lowest-income householders up to $6$ for the highest-income individuals. Examples of mobility graphs of three of these socioeconomic groups are displayed in Figure~\ref{fig:estratos-sociales}.

\begin{figure}[t!]
\centering
\includegraphics[width=0.48\textwidth]{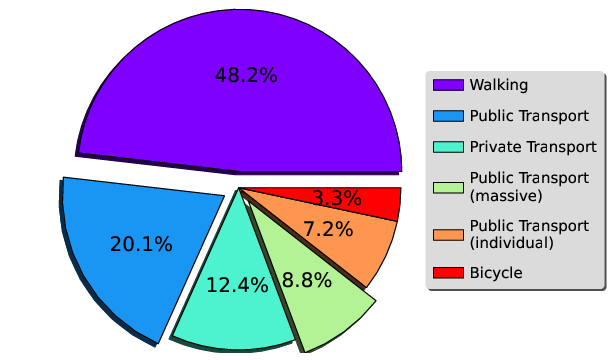}
\includegraphics[width=0.48\textwidth]{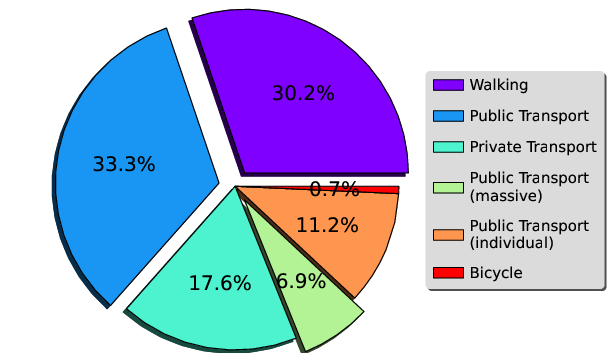}
\caption{Usage of the six transportation modes for the case of Bogot\'a (left) and Medell\'in (right).}
\label{fig:pie-modos}
\end{figure}

\begin{figure}[b!]
\centering
\includegraphics[width=0.45\textwidth]{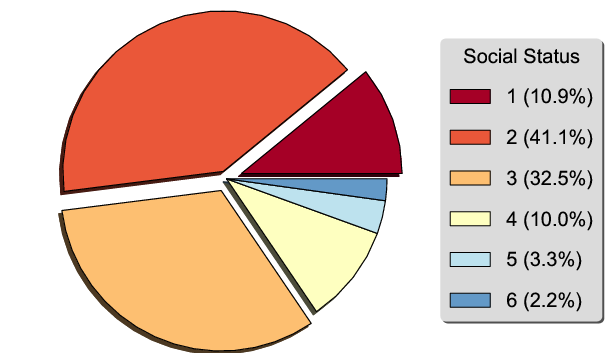}
\includegraphics[width=0.45\textwidth]{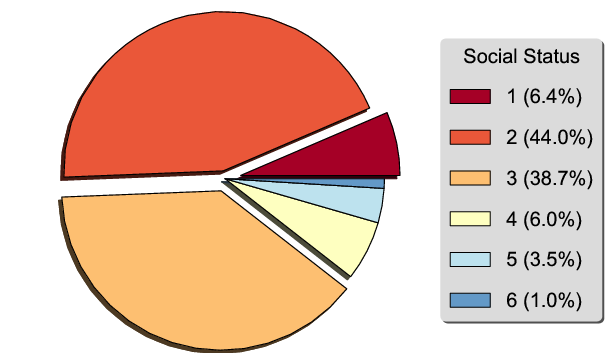}
\caption{Socioeconomic distribution of the population for the cities of Bogot\'a (left) and Medell\'in (right). SES ranging from $1$ (poorest) to $6$ (richest) is assigned to every traveler according to his/her economic wealth.}
\label{fig:pie-statoecon}
\end{figure}

As introduced above, the aim of our work is to study the different means of transportations coexisting in the urban mobility as a multiplex mobility network. Since the surveys contained a number of different transportation means ($25$ in Bogot\'a and $17$ in Medell\'in) we grouped these transportation modes into $6$ different categories. In particular: {\em(i)} pedestrian (walking), {\em (ii)} public transport, {\em (iii)} private transport ({\em e.g.} car or motorbike), {\em (iv)} public massive transport ({\em e.g.} metro), {\em (v)} public individual transport ({\em e.g.} taxi), and {\em (vi)} bicycle. The usage of each transportation group is displayed in Figure~\ref{fig:pie-modos}. Surprisingly, being the same classification for both datasets, we notice that the usage of each group is not the same in both cities. This is due to several factors, such as the different morphology of the cities and the differences in their urban development and planning. Regarding the socioeconomic composition of the population we report in Figure~\ref{fig:pie-statoecon} the partition of the samples used in the surveys of Bogot\'a and Medell\'in in agreement with the real socioeconomic distribution of both cities, being the majority of the population in SES 2 and 3. 

\begin{figure}[t!]
\centering
\includegraphics[height=0.62\textwidth,angle=0]{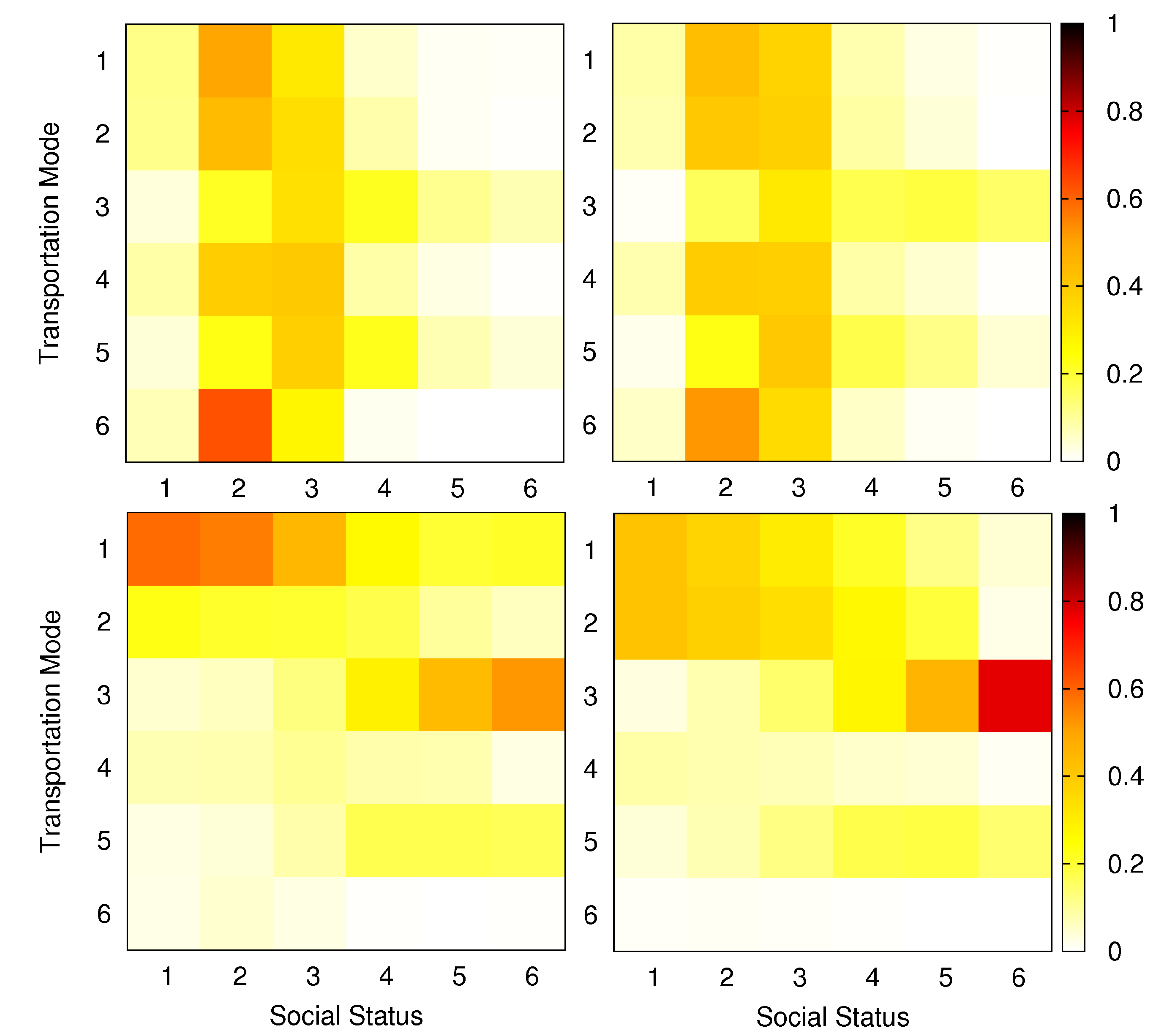}
\caption{Distribution of usage of transportation modes by the $6$ SES. The matrices in the left (right) correspond to the city of Bogot\'a (Medell\'{\i}n). The matrices in the top (bottom) are normalized as $\sum_{s}M_{ts}=1$ ($\sum_{t} M_{ts}=1$). Transportation modes, from $1$ to $6$, are ordered according to Figure \ref{fig:pie-modos}.}
\label{fig:matrix}
\end{figure}

Our goal in the next sections is to use the mobility and socioeconomic data provided by these surveys, to explain how the SES of individuals affect the mobility patterns. To illustrate how the different SES make use of the available transportation modes we show in Figure \ref{fig:matrix} two different mobility matrices, ${\bf M}$. The first type of matrix (top) shows how the usage of a transportation mean is partitioned into the different strata, whereas the second one shows how individuals of a particular SES use the different available modes. In both cases, the latter information is casted in $6\times 6$ matrices which entries $M_{t,s}$ corresponds to, in the first case, the fraction of trips that individuals  belonging to SES $s$ perform using transportation mode $t$. In its turn, in the second case, the entry $M_{t,s}$ accounts for the probability that a trip of an individual of SES $s$ is performed using mean $t$. This information is shown for both cities, Bogot\'a (left) and Medell\'{\i}n (right).

At first sight, comparing those matrices in the left (Bogot\'a) and those in the right (Medell\'{\i}n), both cities display roughly the same usage patterns. In particular, concerning the matrices in the top we can observe that the usage of modes follows two different patterns depending on the precise transportation mean. For instance, modes $3$ (private transport) and $5$ (public individual transport) accumulate users from SES $2$ to $6$, whereas modes $4$ (public massive transport) and $6$ (bicycle) are mainly used by individuals with SES $2$ and $3$.

The second group of matrices (bottom) show, both for Bogot\'a and Medell\'{\i}n, that although multimodality is somehow present for individuals from SES $3$ and $4$, there is a high tendency to concentrate the trips around few transportation means. In fact, this concentration makes clear the socioeconomic differences according to the means selected: poor individuals from SES $1$ and $2$ concentrate their trips using modes $1$ (pedestrian) and $2$ (public transport), which are the cheapest ones, while those belonging to SES $5$ and $6$, mostly use means $3$ (private transport) and $5$ (public individual transport), that represent the most expensive ones. Thus, in the first case the concentration of trips around means $1$ and $2$ is due to the segregation of SES $1$ and $2$ towards cheap means whereas individuals from SES $5$ and $6$ can select their means according to their commodity. Thus, the mobility patterns in both cities show a clear transition segregation-multimodality-selection when going from the poorest to the richest.  
%
%

\section{The Adiabatic Projection of a Multiplex}
\label{sec:methods} 

From the analysis of the mobility matrices in Figure~\ref{fig:matrix}, it becomes clear that the usage of the different transportation modes depends strongly of the status of the individuals. These results demand the analysis of how the different transportation modes are associated forming a \emph{mobility multiplex network} (MMN) for each social status. 
In this section we present the {\em Adiabatic Projection} (AP) technique used to characterize MMN and the structural quantities under study. 

Following the formalism introduced by Battiston \etal ~\cite{battiston-pre-2014}, we consider the MMN of a social status $s$ as a system composed of $N$ nodes and $M=6$ layers. As explained before, nodes correspond to the different urban areas in a city. Layers, instead, represent different transportation modes. Keeping in mind such setup, and particularizing in the mobility multiplex of a given social status $s$, it is possible to associate to each layer $\alpha$ ($\alpha=1,\ldots,M$) a graph ${\mathcal G}^{[s,\alpha]} (N, \mathcal{E}^{[s,\alpha]})$ described by an adjacency matrix ${\bf A}^{[s,\alpha]}$ whose entries are defined as $a_{ij}^{[s,\alpha]}= 1$ if zones $i$ and $j$ are connected by (at least) a trip of an individual from status $s$ using transportation mode $\alpha$. Under this formalism, the MMN of a social status $s$ is fully described by the so-called \emph{vector of adjacency matrices} $\mathbf{A}^s$ given by:
\begin{equation}
\label{eq:adj_vec_multi}
\mathbf{A}^s = \left\{ {\bf A}^{[s,1]}, \ldots , {\bf  A}^{[s,M]} \right\}\,.
\end{equation}

Once having introduced the basic notation characterizing each of the MMN, we describe the AP procedure used to study the coexistence of several interaction (here transportation) modes in a multiplex network. The technique relies in merging together a subset $V(m)$ containing $m\leq M$ layers into a single (monolayer) graph ${\mathcal G}^{[s,V(m)]}(N,\mathcal{E}^{[s,V(m)]})$ where:
\begin{equation}
{\mathcal{E}}^{[s,V(m)]}=\bigcup_{\alpha\in V(m)}  \mathcal{E}^{[s,\alpha]}\;.
\end{equation}
Therefore, the network ${\mathcal G}^{[s,V(m)]}$ is obtained by projecting all the layers contained in $V(m)$ onto a single one and by converting the multiple  links (those existing in several layers in $V(m)$) into single ones. In this way, the topology of the resulting projected network is described by the \emph{projected adjacency matrix} ${\bf A}^{[s,V(m)]}$ defined as:
\begin{equation}
a^{[s,V(m)]}_{ij} = %
\begin{cases}
1& \text{ if } \sum_{\alpha\in V(m)} \; a_{ij}^{[s,\alpha]} > 0\\
0& \text{ if } \sum_{\alpha\in V(m)} \; a_{ij}^{[s,\alpha]} = 0 \;.
\end{cases}
\end{equation}

The purpose of the AP of the layers of a multiplex is to analyze the evolution of some topological quantities when passing from single layers to the projected network resulting from merging all the $M$ layers of the multiplex. Thus, the approach, introduced in \cite{cardillo-scirep-2013} to study the {\em European Air Transportation Multiplex}, consists in varying the number of layers contained in the subset $V(m)$ from $m=1$ to $m=M$. It is important to notice that the AP method (as introduced in \cite{cardillo-scirep-2013}) considers, for each value $m$, the set ${\mathcal V}(m)$ containing all the possible subsets $V(m)$ comprising $m$ layers. In this way, given a topological quantity $x$, one evaluates $x$ in each projected graph ${\mathcal G}^{[s,V(m)]}$ derived from each subset $V(m)$ contained in ${\mathcal V}(m)$ and average the values obtained over all the resulting graph. Thus, given $m$, the average value of $x$ in ${\mathcal V}(m)$ reads: 
\begin{equation}
\langle x\rangle(m)=\frac{m! (M-m)!}{M!}\sum_{V(m)\in {\mathcal V(m)}} x({\mathcal G}^{[s,V(m)]})\;.
\label{eq:xmed}
\end{equation}
Note that, although for $m=1$ there are $M$ possible subsets in ${\mathcal V}(1)$ whereas for $m=M$ there is only $1$ subset $V(M)$, for a general value  $m$ the cardinal of  ${\mathcal V}(M)$ can be extremely large. Thus, the AP technique demands a computationally expensive statistical treatment to cover all the possible layer combinations included in the sum of equation \ref{eq:xmed} when the number of layers $M$ is large enough.

Here, instead, we get rid off the statistics over the sets ${\mathcal V}(m)$. In particular, based on the details contained in the dataset, we make use of the information about the usage of each transportation mode $\alpha$ by each SES $s$ so that, for a certain value of $m$, we consider the projected graph ${\mathcal G}^{[s,V(m)]}$ constructed by merging the $m$ most used transportation modes (layers) by SES $s$. 
In this way, for each value of $m$, ${\mathcal V}(m)$ contains one single subset $V(m)$ and thus we will denote each projected graphs as ${\mathcal G}^{[s,m]}$ and its associated adjacency matrix as ${\bf A}^{[s,m]}$.
Apart from the computational simplification of this variant of the AP technique, the new path from $m=1$ to $m=M$ informs about how the individuals of a particular SES are benefited by adding transportation modes to their trips allowing to distinguish between strata displaying either segregation or selection of modes and those socioeconomic compartments showing multimodality.

The topological quantities studied under the AP technique cover traditional structural measures, used in simple networks, and others that take into account the layer structure of a multiplex. In particular, for each graph  ${\mathcal G}^{[s,m]}$ we will study the following usual properties:
\begin{itemize}

\item {\em The size, $S$, of the giant component and the number of components, $n_c$}. It is important to note that $S$ is normalized to be $0\leq S\leq 1$, so that $S=1$ when the $N$ nodes in the network take part of a unique component. In addition, to compute $n_c$ we have considered that isolated nodes do not constitute a component so that components contributing to $n_c$ are those of size equal or larger than $2$.

\item {\em The average path length, $L$}. As usual, $L$ is the average length of the shortest paths among all the couples of nodes in the network. Since the networks under study are highly disconnected, especially for small values of $m$, we have adopted the typical way out to avoid divergences in $L$, {\em i.e.}, to consider only the nodes in the giant component. 

\item{\em The average degree, $\langle k\rangle$}. Again, in order to compute the average number of connections of the nodes we have excluded isolated nodes.

\item {\em The clustering coefficient, $C$}. As usual, the clustering coefficient shows the probability that two nodes $i$ and $j$ having a common neighbor $l$ are also connected. In this case also, isolated nodes do not contribute to clustering.
\end{itemize}

The above measures are those traditionally used for characterizing simple (single-layer) networks. However, there also exist measures that are specifically designed for multiplex networks (see the recent reviews \cite{boccaletti-rev-2014,kivela-rev-2014}). This is the case of the \emph{Overlap}, $O$. The overlap quantifies the redundancy of links between layers, {\em i.e.}, the fact that a link between two given nodes $i$ and $j$ is present in several layers. In our multiplex networks the existence of a large overlap would imply a large tendency of the individuals (belonging to the same SES) of using different means of transportation for connecting the same urban areas $i$ and $j$.  In the recent years, several overlap measures have been proposed \cite{barigozzi-pre-2010, bianconi-pre-2013, kapferer-1969,parshani-epl-2010, battiston-pre-2014}. Here, for a given value of the number of projected layers $m$, the overlap of the resulting graph ${\mathcal G}^{[s,m]}$ is measured via two different quantities, namely:
\begin{eqnarray}
O_1  &=& \dfrac{W - K}{K}\,,
\label{overlap1}
\\
O_2  &=& \dfrac{D}{K}\,.
\label{overlap2}
\end{eqnarray}
Where $K$ is the number of links in the aggregate graph ${\mathcal G}^{[s,m]}$, $W$ is the total sum of the links in each of the $m$ layers merged in ${\mathcal G}^{[s,m]}$, and $D$ is the number of redundant links in the set of $m$ layers. 
We can express these quantities, $W$, $K$ and $D$, making use of the adjacency matrices associated to each layer, ${\mathcal G}^{[s,\alpha]}$, and that of the projection of the $m$ most used layers, ${\mathcal G}^{[s,m]}$, as:
\begin{eqnarray}
W &=& \sum_{\alpha=1}^{m} \; \sum_{i,j=1}^{N}a^{s,\alpha}_{ij}\,,
\\
K &=& \sum_{i,j=1}^N a_{ij}^{s,m}\,,
\\
D&=&\Theta\left( \sum_{\alpha=1}^m a^{s,\alpha}_{i,j}-2 \right)\;,
\end{eqnarray}
where, in the last equation, $\Theta(z)$ is the step function defined as $\Theta(z)=0$ for $z<0$ and $\Theta(z)=1$ otherwise.

\section{Results}
\label{sec:results}

In this section, and relying on the the AP technique of the MMN, our aim is to unveil the mobility patterns associated to the use of the transportation modes of each SES and, moreover, to monitor how the different patterns present in the transportation layers are combined into their corresponding mobility networks.

\begin{figure}[t!]
\centering
\includegraphics[width=0.97\textwidth]{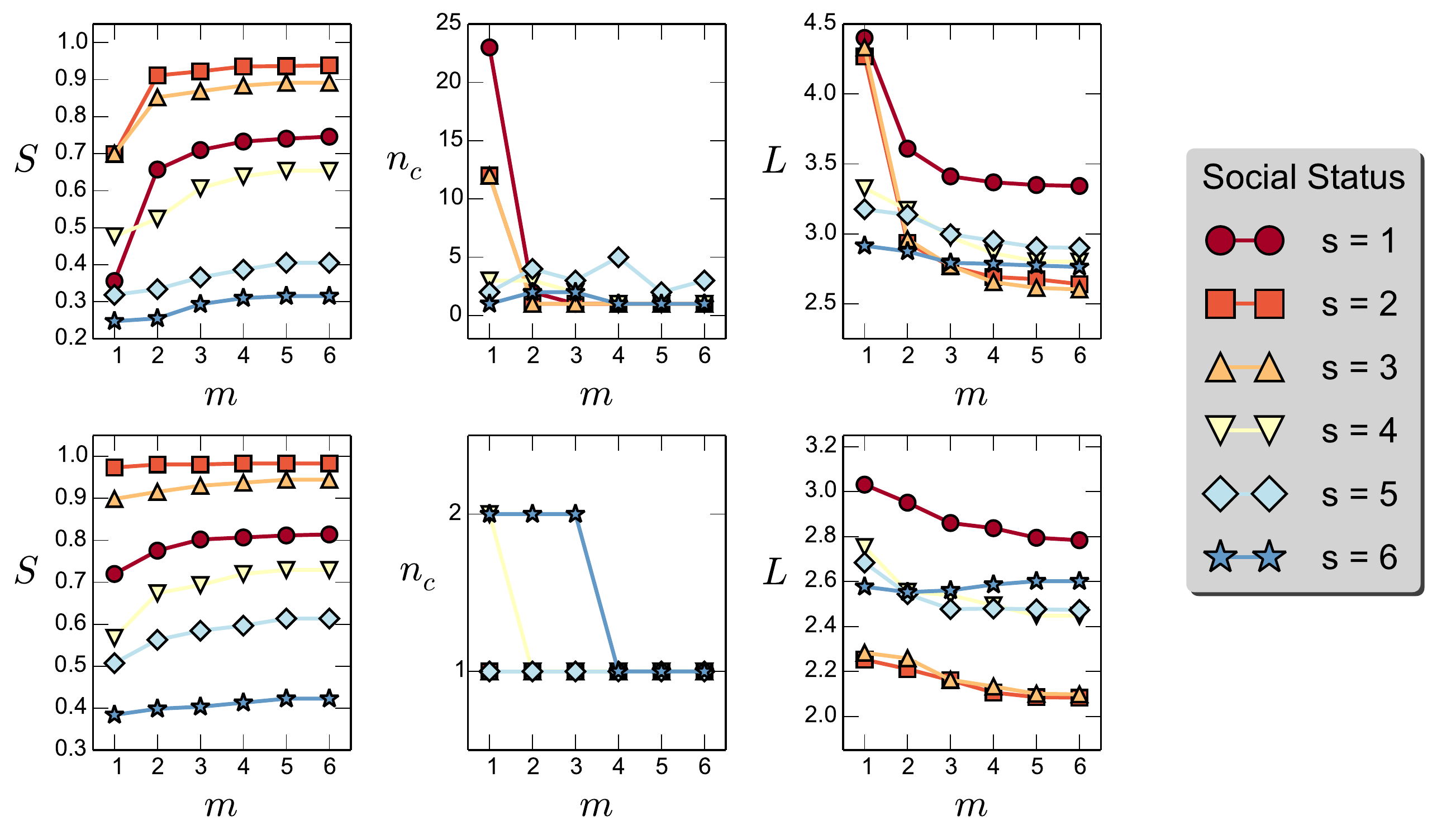}%
\caption{Adiabatic evolution of the size of the giant component $S(m)$ (left panels), number of components $n_c(m)$ (central panels), and average path length $L$ (right panels). The top panels are for the city of Bogot\'a, while those in the bottom refer to Medell\'in.  Each curve corresponds to a different SES. Namely: 1 ($\bullet$), 2 ($\blacksquare$), 3 ($\blacktriangle$), 4 ($\blacktriangledown$), 5 ($\blacklozenge$) and 6 ($\bigstar$).}
\label{fig:res-components}
\end{figure}

We start by studying how the combination of different transportation modes cover the different urban areas. This view can be explained as a percolation process driven by the addition of network layers  (instead of nodes or links as in the traditional percolation contexts). To this aim, we focus on the evolution with $m$ of the size of the giant connected component, $S(m)$, as well as the evolution for the number of components $n_c(m)$, and that of average path length of the giant component, $L(m)$. In Figure~\ref{fig:res-components}, we show these evolutions for each of the $6$ SES for the cases of Bogot\'a (top) and Medell\'{\i}n (bottom). 

The adiabatic evolution of the giant component $S(m)$ shows that both cities behave in a similar way so that the the different evolution $S(m)$ for the SES follows the same hierarchy. In particular, SES $2$ and $3$ reach to cover almost all the urban mobility zones of the cities. On the other hand, the coverage of SES $6$ in both cities and also $5$ in Bogot\'a are well below the $50\%$ of the  
zones. The main difference between the two cities shows up by looking at the rate $S(m)$ increases. While for Medell\'{\i}n the rate of change is very small for all the SES, in Bogot\'a, SES $1$, $2$ and $3$ need to merge at least two different transportation layers in order to achieve the $80\%$ of their corresponding coverage. This result is the fingerprint of the segregation of these poor SES observed combined with the effect of the smaller sample size of the Bogot\'a survey (as compared to that of Medell\'{\i}n) that makes difficult to capture weak connections between urban mobility zones. These fact seems to affect more poor SES due again to their spatial segregation.  

The evolution of the number of components $n_c(m)$, and the average path length $L(m)$ in the city of Bogot\'a further confirms the effects of the segregation of SES $1$, $2$ and $3$. As observed, the initial ($m=1$) values of both $n_c$ and $L$ are extremely large and they need to merge at least two transportation modes to reach small values of $n_c$ and $L$. This is not the case for SES $4$, $5$ and $6$ for which the evolution is far more smooth. Concerning the final values of $L$ in the city of Bogot\'a, it is remarkable the large steady value reached by SES $1$ as compared to the rest of the population. Thus, even if they can cover a large number of zones the trips connecting them associate in a rather linear way, thus not displaying shortcuts. In its turn, the situation in Medell\'{\i}n concerning the evolution of $n_c(m)$ is not pretty much like to that of Bogot\'a. In fact, in this city the locations of the usual destinations appear to be very clustered, leading to have a system composed of only one component even for $m=1$ for most of the SES. The evolution of $L(m)$ instead is more interesting. As in the case of Bogot\'a, $L$ decreases with $m$ although in a smoother way [as occurred for the evolution of $S(m)$]. Again, it is worth to notice how, as in the case of Bogot\'a, SES $1$ displays a different behavior from the rest of strata.

Summarizing, both the behavior of $S(m)$ and $L(m)$ point out that in both Bogot\'a and (more clearly) Medell\'in the six SES can be regrouped into three mobility compartments related with their wealth. Namely: low (SES $1$), mid-low (SES $2$ and $3$) , and mid-high (SES $4$, $5$ and $6$) compartments. 

\begin{figure}[t!]
\centering
\includegraphics[width=0.87\textwidth]{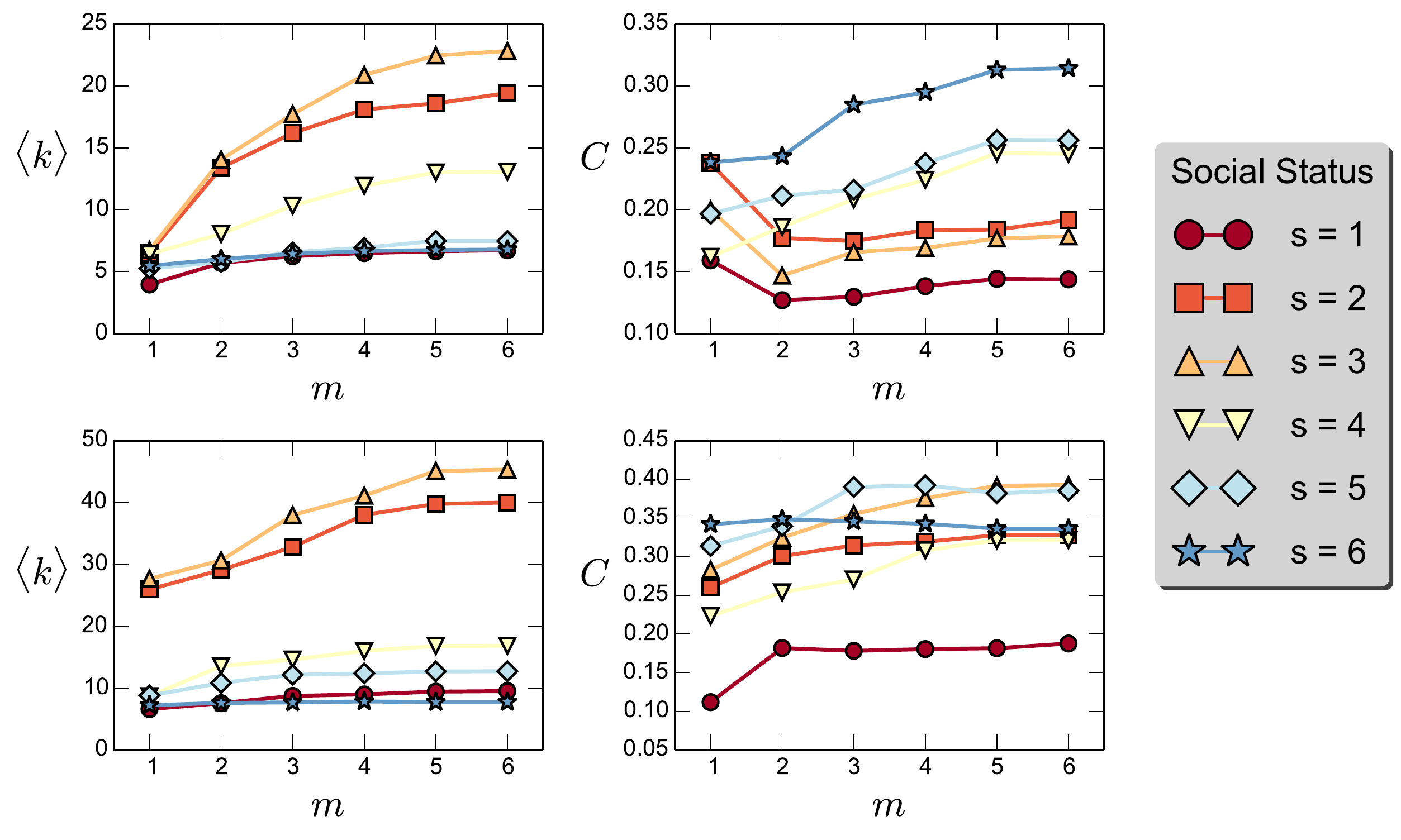}
\caption{Adiabatic evolution of the number of different destinations reached from a place, $\langle k\rangle$ and clustering coefficient, $C$ as a function of the number of layers merged together, $m$. Left panels display the evolution for the average connectivity $\avg{k}(m)$, while those in the right show that of the clustering coefficient $C(m)$. The top panels are for the city of Bogot\'a, while those in the bottom refer to Medell\'in.  Each curve corresponds to a different SES. Namely: 1 ($\bullet$), 2 ($\blacksquare$), 3 ($\blacktriangle$), 4 ($\blacktriangledown$), 5 ($\blacklozenge$) and 6 ($\bigstar$).}
\label{fig:res-connectivity}
\end{figure}

In Figure~\ref{fig:res-connectivity}, we confirm the above compartmentalization by monitoring the evolution for the number of different trips from/to each urban area (here represented as the average degree $\langle k\rangle$ of the nodes) and the role of the various transportation modes in the triadic closure phenomenon, here studied via the clustering coefficient $C$. The evolution $\langle k\rangle(m)$ reveals two clearly distinct behaviors. First, for SES $2$ and $3$ (also $4$ in the city of Bogot\'a) incorporating transportation modes implies to increase the number of origins/destinations, pointing out the genuine multimodal character of these individuals who assign different transportation modes depending on the trip to be performed. On the other hand, for the rest of the SES there is almost no evolution. However, when looking back to the evolution of $S(m)$ in Figure~\ref{fig:res-components}, it is easy to notice that the almost steady behavior of $\langle k\rangle(m)$ for these strata has different roots. While individuals belonging to SES $5$ and $6$  move from/to a limited amount of different places [as displayed by the small values of $S(m)$] using few transportation modes, due to the aforementioned selection mechanism, SES $1$ displays a large coverage. Thus, for SES $1$, the addition of a new transportation layer is mostly devoted to join pairs of disconnected nodes, and thus not used to increment the communication power of zones for which a trip already exists. 
	
The particular way of evolution with $m$ displayed by SES $1$ is also related to the large resulting networks [as displayed by $L(m)$ in Figure~\ref{fig:res-components}] and further confirmed by looking at the evolution of the clustering coefficient, $C(m)$. In both cities the values displayed by SES $1$ are the smallest of the population and it does not show any significant change when increasing $m$. At variance, SES $5$ and $6$ display the largest values for the clustering in both cities, thus confirming again that, in this cases individuals cover a limited and rather fixed number of zones, thus favoring the formation of triadic paths in the aggregated graph. 

\begin{figure}[t!]
\centering
\includegraphics[width=0.87\textwidth]{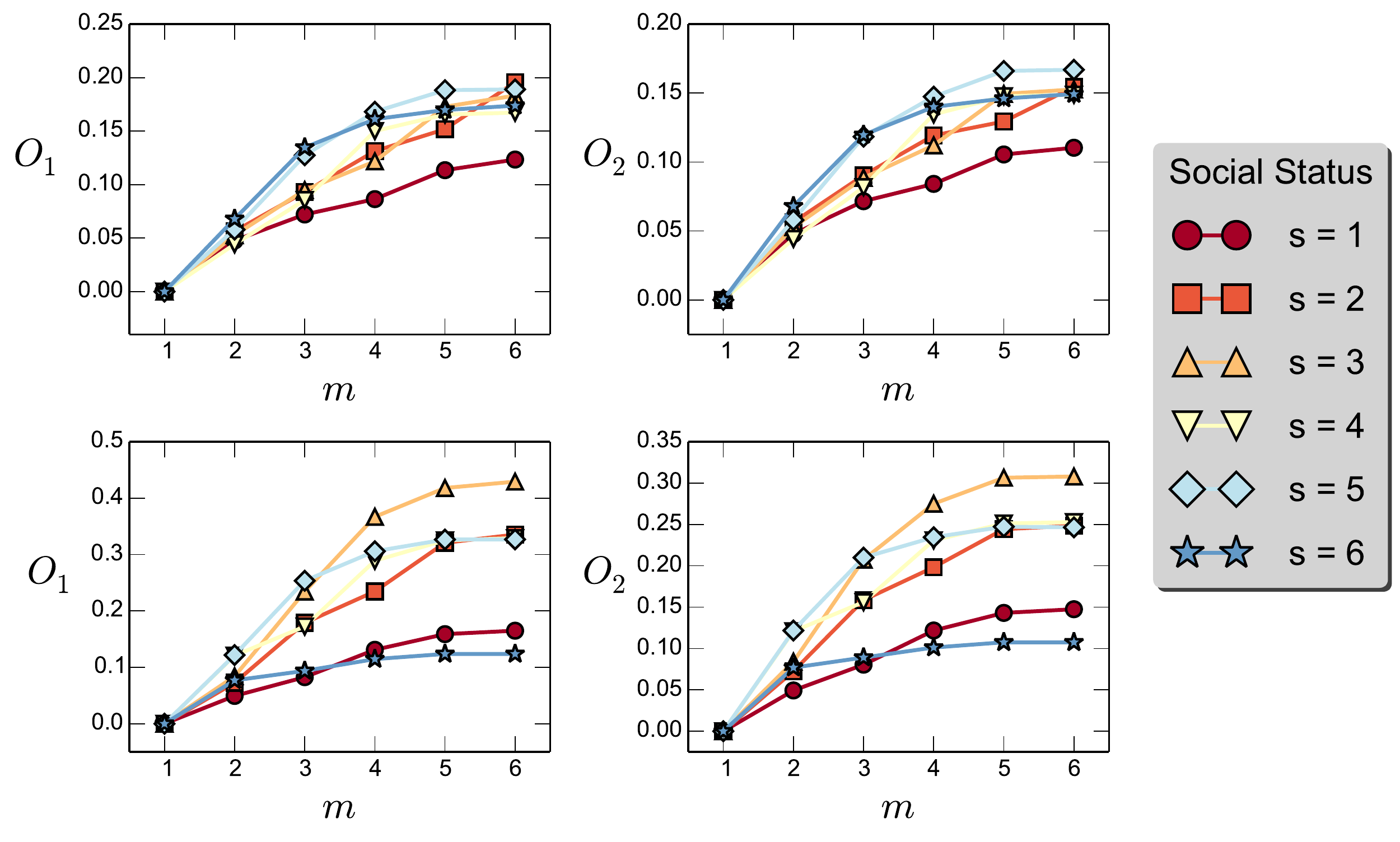}
\caption{Adiabatic evolution of the overlap as a function of the number of layers merged together, $m$. Left panels display the evolution for first definition of overlap, $O_1$, Equation \ref{overlap1}, while those in the right show the second one, $O_2$, Equation \ref{overlap2}.  The top panels are for the city of Bogot\'a, while those in the bottom refer to Medell\'in.  Each curve corresponds to a different SES. Namely: 1 ($\bullet$), 2 ($\blacksquare$), 3 ($\blacktriangle$), 4 ($\blacktriangledown$), 5 ($\blacklozenge$) and 6 ($\bigstar$).}
\label{fig:res-overlap}
\end{figure}

Finally, in Figure \ref{fig:res-overlap} we present the evolution for two measures of the overlap proposed above (see Equations \ref{overlap1} and \ref{overlap2}). Interestingly, for the two cities the two measures present almost the same evolution in terms of the relative growth between the different SES. Considering the definitions of $O_1$ (that takes into account the total amount of degeneration of the links) and $O_2$ (that only count once the redundant links regardless of the number of times they are repeated) it is clear that $O_1\geq O_2$. However, the similar trends observed and the small difference in the values attained by $O_1$ and $O_2$ point out that all of the SES do not tend to accumulate more than two overlapping links. Concerning the differences in the increase rates of $O_1$ and $O_2$ between SES we observe that, in both cities, individuals belonging to SES $1$ tend to avoid overlapping, in agreement with the way (discussed before) SES $1$ tend to increase the size of the giant component. Importantly, for the city of Bogot\'a the trends observed in both $O_1(m)$ and $O_2(m)$ seem to reproduce the three mobility compartments discussed above ($1$, $2-3$ and $4-5-6$) for both cities. However, the results for the city of Medell\'{\i}n are completely in disagreement with these compartments since, for instance, SES $6$ display small overlapping tendency (being similar to that of SES $1$) in contrast to the large tendency of SES $4$ and $5$. 

\section{Conclusions}
\label{sec:conclusions}

We have presented a dataset about the human mobility in urban areas with two ingredients of utmost interest: the information about the multimodal nature of the trips, and the \textit{socioeconomic status} of the individuals. The first ingredient has allowed us to tackle the analysis of the mobility patterns using the multiplex framework, which has attracted many attention lately. On the other hand, the information about SES provides with a novel ingredient that has been ignored up to date in the studies about human mobility. Exploiting these two ingredients, the aim of this manuscript has been to describe how the different socioeconomic compartments make use and combine the different transportation layers.

We have analyzed the mobility multiplex networks of each SES by studying how different structural descriptors evolve as network layers (the transportation modes) are merged. This procedure, called adiabatic projection, starts from the network of the trips performed by means of the most used  transportation mode and subsequently adds the layers corresponding to the other means in descending order of usage. 

The main result of our work is the classification of the $6$ SES into three compartments according to their behavior. Namely, in a first group we have SES $1$ and $2$, the poorest ones, whose behavior is characterized by the segregation, {\em i.e.}, the usage of few and cheap transportation modes to cover a large fraction of the urban areas in a rather sparse way. The second compartment is composed of SES $3$ and $4$, having a genuine multimodal pattern and covering almost the total number of urban zones. Finally, the elite compartment composed of SES $5$ and $6$, is characterized by a selection of costly modes for performing the trips that, in their turn, display a very small coverage in terms of the urban areas reached although the connectivity within these zones turns to be rather dense. 

The unveiled differences in the organization of the mobility multiplex networks according to SES demands the inclusion of this novel ingredient in the studies about human mobility and intrinsically related processes. As an example, it would be of interest to incorporate the presence of socioeconomic differences when studying the development of contagion processes in urban areas. We hope that our work will motivate more studies in this direction.

\begin{acknowledgement}
We acknowledge financial support from the European Commission through FET IP projects MULTIPLEX (Grant No. 317532) and PLEXMATH (Grant No. 317614), from the Spanish MINECO under projects FIS2011-25167 and FIS2012-38266-C02-01, from the Comunidad de Arag\'on (Grupo FENOL), and from the  Universidad Nacional de Colombia under grants HERMES 19010 and HERMES 16007. JGG is supported by the Spanish MINECO through the Ram\'on y Cajal program. AC acknowledge the financial support of SNSF through the project CRSII2\_147609. We thank \textit{Area Metropolitana del Valle de Aburr\'a}, in Medell\'in, and \textit{Secretar\'ia Distrital de Movilidad}, in Bogot\'a, for the Origin-Destination Surveys Datasets. 
\end{acknowledgement}


\end{document}